\documentclass[
]{ceurart}

\usepackage{listings}
\usepackage{enumitem}
\begin{document}

\copyrightyear{2024}
% CC-BY is default license.
\copyrightclause{Copyright for this paper by its authors. Use permitted under Creative Commons License Attribution 4.0 International (CC BY 4.0).}

\conference{ReNeuIR 2024 (at SIGIR 2024) -- 3rd Workshop on Reaching Efficiency in Neural Information Retrieval, 18 July, 2024, Washington D.C, USA}

\title{Efficient course recommendations with T5-based ranking and summarization}

% \tnotetext[1]{You can use this document as the template for preparing your publication. We recommend using the latest version of the ceurart style.}

\author[1,2]{Thijmen Bijl}[%
orcid=0009-0000-9550-2502,
%email=kulyabov-ds@rudn.ru,
%url=https://yamadharma.github.io/,
]
%\cormark[1]
\address[1]{Randstad, the Netherlands}
\address[2]{Leiden Institute of Advanced Computer Science, Leiden University, the Netherlands}

\author[1]{Niels van Weeren}[%
% orcid=0000-0001-7116-9338,
% email=i.tiddi@vu.nl,
% url=https://kmitd.github.io/ilaria/,
]
\author[2]{Suzan Verberne}[%
orcid=0000-0002-9609-9505,
email=s.verberne@liacs.leidenuniv.nl,
% url=http://conceptbase.sourceforge.net/mjf/,
]

%\cortext[1]{Corresponding author.}
%\fntext[1]{These authors contributed equally.}

\begin{abstract}
%1. What did we do
In this paper, we implement and evaluate a two-stage retrieval pipeline for a course recommender system that ranks courses for skill--occupation pairs.
%2. Why did we do it
The in-production recommender system BrightFit provides course recommendations from multiple sources. Some of the course descriptions are long and noisy, while retrieval and ranking in an online system have to be highly efficient. 
%3. How did we do it
We developed a two-step retrieval pipeline with RankT5 finetuned on MSMARCO as re-ranker. We compare two summarizers for course descriptions: a LongT5 model that we finetuned for the task, and a generative LLM (Vicuna) with in-context learning. We experiment with quantization to reduce the size of the ranking model and increase inference speed. We evaluate our rankers on two newly labelled datasets, with an A/B test, and with a user questionnaire.
%4. What did we find
On the two labelled datasets, our proposed two-stage ranking with automatic summarization achieves a substantial improvement over the in-production (BM25) ranker: nDCG@10 scores improve from 0.482 to 0.684 and from 0.447 to 0.844 on the two datasets. We also achieve a 40\% speed-up by using a quantized version of RankT5. The improved quality of the ranking was confirmed by the questionnaire completed by 29 respondents, but not by the A/B test. In the A/B test, a higher clickthrough rate was observed for the BM25-ranking than for the proposed two-stage retrieval. 
%5. What do we think it means
We conclude that T5-based re-ranking and summarization for online course recommendation can obtain much better effectiveness than single-step lexical retrieval, and that quantization has a large effect on RankT5. In the online evaluation, however, other factors than relevance play a role (such as speed and interpretability of the retrieval results), as well as individual preferences.
\end{abstract}


\maketitle

\section{Introduction}
With a rapidly changing labour market, it is becoming more important for people to adapt to these changing demands in order to remain relevant to employers. The topic of \textit{skilling} (learning new skills) plays an important role in the current labour market, both for employers and employees. For employers, it is in their best interest to let their employees learn new skills in order to remain competitive. For the employees, it is important to keep learning new skills in order to grow in their current job or get better career opportunities at other organizations. 

In the ``Future of Jobs" report by the World Economic Forum (WEF)~\cite{WEF-Future-of-jobs}, it is estimated that 44\% of the in-demand skills will change between 2023 and 2028. Another survey among 2000 HR professionals and employees indicates that there is much interest in re- and up-skilling~\cite{risesmart-survey}. However, the survey also finds that while both employees and employers see the benefits of learning new skills, they struggle with finding and identifying good skilling opportunities. %Both employees and employers indicate that they would benefit greatly from professional guidance to help them identify what skills to learn along with help to find proper resources to learn these skills.

The online recommender system \textit{BrightFit}, a system developed by Randstad Risesmart\footnote{\url{https://go.randstadrisesmart.com/BrightFit}}, 
aims to help users explore future potential roles by looking at their current skill set and determining the skill gap to the next job or role they aspire to get. %The users of BrightFit are either people who have recently lost (or about to loose) their jobs and need to adapt their skills to find a new job or they are people who are looking to take the next step in their career. 
%A big part of getting to their new desired role is to close the skill gap between a user's current skill set and the required skill set for their new potential role. 
For this, BrightFit recommends online courses from several mainstream course providers to help a user learn a particular skill. 

In this paper, we approach the problem of recommending courses as an information retrieval (IR) problem. This alleviates the cold start problem that collaborative or content-based filtering approaches suffer from: both require historical user data to predict the users' preferences. The users of BrightFit are typically short-term users for whom historical profile data is not available. Thus, we are not recommending courses based on previous user activities, but based on a combination of the user's job profile and the skill they want to learn.  
A query consisting of the skill and the job profile is automatically constructed and issued to the index with course descriptions. %based on the user's interactions with the BrightFit application. 

The current ranker in BrightFit's backend is BM25-based. We propose a two-stage retrieval method that leverages the power of transformer-based models while keeping the time to generate recommendations reasonable. We first evaluate two first-stage retrievers (the dense retriever GTR and BM25), followed by the re-ranker RankT5~\cite{zhuang2022rankt5}.

Since we do not have any human-labelled training data, the ranking is performed in a zero-shot setting. For evaluation, we compile and label 2 datasets with relevance assessments on three levels, for over 2500 query--document pairs. %

Course retrieval comes with an additional challenge that the course descriptions are often long and contain irrelevant information, making it more difficult to efficiently retrieve the correct courses. We therefore evaluate summarization methods in our retrieval pipeline to both shorten and de-noise the course descriptions. We investigate if we can we improve the ranking by summarizing the course descriptions. 
We compare two summarization models: LongT5 (an encoder-decoder model), which we finetune on pairs of long and short course descriptions, and Vicuna (a decoder-only LLM), with a zero-shot instruction prompt.

Since we are providing recommendations to real users in real-time, inference of the models has to be fast: BrightFit has a few seconds at most to provide the recommendations. 
We therefore investigate model quantization (compressing the parameters of the RankT5 model)~\cite{quantization-theory} as a way to increase inference speed and decrease storage/memory requirements. This can come at the cost of reduced model performance since we store the weights in lower precision. We evaluate quantization of our RankT5 model without compromising the quality of the recommendations too much.

Lastly, we evaluate how the improvements in offline datasets translate to the user's perception of the recommendations. Prior work has indicated the importance of user studies for the evaluation of recommender systems~\cite{martin2009recsyskeynote,Knijnenburg2015}. 
We use an A/B test and a user questionnaire to evaluate whether our changes to the recommendations in BrightFit make a positive impact on the usage of the system.

In summary, we make the following contributions:
\begin{itemize}
    \item We create two labelled datasets for evaluation of course retrieval for skills and occupations.\footnote{We released our data through \url{https://github.com/tbijl/course_ranking_data} Note that the documents are listed in the form of URLs in their course provider domain.}
    \item We show that automatic summarization of the course descriptions improves the effectiveness of the course ranking.
    \item We show that quantization of RankT5 does not reduce the effectiveness, while substantially improving the inference speed of the ranker.
    \item With an online A/B test with real users and explicit user feedback in a questionnaire, we show that evaluation of course recommendation is challenging and that other factors than relevance play a role in user preferences. 
\end{itemize}

\section{Related work}
\paragraph{Course recommendations}
The previous work on course recommendations is largely based on user profiles. Imran et al.~\cite{Imran2015} ask users about their prior knowledge to establish a user profile. This profile is extended based on user behaviour in the recommender system. In the context of MOOCs (Massive open online courses), Jing and Tang~\cite{jing2017guess} identify the challenges of user behavior modeling for course recommendation. They also argue that students might participate in courses for different reasons, and therefore collaborative filtering methods are not successful for course recommendation. They propose the use of historical user data to build an interest profile. For our use case, we do not have user behaviour history for the majority of the users, because BrightFit is used as a short-term tool and users have a large variety of occupations and skillsets. We therefore recommend courses for a specific skill, not based on a user profile. %Additionally, we want to be able to make good recommendations to the user from the first time they start using the system, which means that we cannot rely on any previous interactions by the user.

Because skills can be organized in structured databases, ontology-based approaches are commonly used in prior work on course retrieval and recommendation~\cite{ahmed2010ontology,GEORGE2019103642}. %Creating and using ontologies in recommender systems requires a lot of knowledge engineering and time. 
One approach is to map learners and learning materials to topics in an ontology, and then recommend courses for topics to the learners using the relations between items in the ontology~\cite{Shishehchi,lit0}. 
%To overcome this issue of having to map items manually to items in an ontology, Korel et al.~\cite{computers12010014} propose the use of BERT as a classification tool to map resources to items in an ontology allowing for easy search based on the items in the ontology. The advantage of this method is that it is much more scaleable than manually assigning ontology items to each learning object. But the automatic ontology linking task is a challenge. %They found that the method has difficulty in creating similar embeddings for concepts in the ontology as for the content that should be related to those concepts. 
%Additionally, they don't have any labelled ground truth dataset to evaluate their methods fully.
The alternative is to start with a skills ontology and then collect publicly available resources as educational content for each skill in the ontology~\cite{Tavakoli2022}. %They perform quality assessment and metadata analysis on the collected content. Additionally, they combine their resources with a user profile which contains preferences for certain types of content to make recommendations that fit the user's preference. 
Ontology-based approaches can be effective for finding and recommending courses that are related to a certain topic, however, they come with the drawback of knowledge engineering to map all courses to topics in the ontology. This makes these approaches unfeasible to use in contexts with a pre-defined large collection of unstructured course descriptions, like in BrightFit. In this setting, a text retrieval approach is more fit.

\paragraph{Quantization of ranking models} There is only limited prior work on the use of quantization methods to make ranking models more efficient. Quantization has been applied in prior work to the encoding of token embeddings ~\cite{yang2022compact} %: ``proposes contextual quantization of token embeddings by decoupling document specific and document-independent ranking contributions during codebook-based compression''
and to efficient storage of decision trees in learning-to-rank ensembles~\cite{gil2022ensemble}. %: ``investigate binning and quantization techniques to reduce the memory occupation of ensemble models for learning to rank.'' Decision trees, ``We apply quantization to the leaf values of each tree of the ensemble, so as to further lower the memory footprint. ''
Recent work has addressed quantization for the efficient coding of the document IDs in generative IR~\cite{zeng2023scalable,pradeep2023does}. In one prior work \cite{lassance2023static}, quantization is mentioned as a suggestion to further improve the efficiency of SPLADE (sparse neural retrieval). To the best of our knowledge, our paper is the first to apply quantization to RankT5 or other transformer-encoder ranking models.

% \paragraph{Recommender system evaluation}
% In a keynote by Martin Francisco at the 2009 ACM Recommender Systems Conference (RecSys)~\cite{martin2009recsyskeynote} he discusses the importance of evaluating recommender systems using real user experiments. He argues that there are a lot of factors that play a role in the user preference of one recommendation over others. He expands on this by saying that other interactive components of the recommender system can account for up to 50\% of the success of a recommender system, while the actual algorithm used for recommendations can account for only as little as 5\%. 

% A survey by Knijnenburg et al.~\cite{Knijnenburg2015} gives an overview of different methods and important considerations when looking to perform user-based evaluation of a recommender system. One important aspect that is mentioned is that the optimal test plan for evaluating recommender systems involves both an A/B test and a questionnaire or survey. A/B tests are better at uncovering interesting effects of how the recommendation system behaves and integrates with a system. An additional survey is useful to get more explicit feedback on choices made by users. This can help to get better insight into observed user behaviour from the A/B test. A survey enables us to get an idea about what the users are thinking when processing the recommendations.  

\section{Data collection}
\subsection{Course data sources}

BrightFit offers courses from four different online learning platforms.  %including edX and Udemy.  
Table~\ref{tab:course-counts} shows an overview of the total number of courses per provider. 
\textit{Udemy} is an online learning platform that offers video courses about a large variety of topics. Udemy is a marketplace where users can create and sell their own courses or buy and take courses created by other users. %Udemy has a selection of both paid and free courses available. The model of offering community-created courses has its pros and cons. The main advantage is that the courses are created by a large community of users, which results in a large offering of courses on a large variety of topics. The main disadvantage of having community-created courses is that there is little to no guarantee of the quality of the courses that are offered which can make it harder to recommend good courses to users. %To get an indication of the quality of the courses that BrightFit collects from Udemy, we show an overview of the rating data of the courses that are collected in Figure~\ref{fig:udemy-distribution}. From the distribution we see that most courses have a rating higher than 4 stars, however, in the review count distribution plot, we see that overall these ratings are based on a low number of reviews.
We retrieved course metadata from Udemy using the public Udemy affiliate API.\footnote{\url{https://www.udemy.com/developers/affiliate/}} This API returns general information about the courses that Udemy provides. %The fields that are most relevant for our use case are described in Table~\ref{tab:udemy-api-fields}.
In total, Udemy has over $210,000$ courses available on its website. %Due to an API limitation, we are not able to get data on all 210,000 Udemy courses. In this thesis, 
We only use English courses. %At the time of writing BrightFit has a total of $54,348$ English courses provided by Udemy in its catalogue. 
\textit{Udemy Business} is similar to Udemy. It is a subscription service offered by Udemy to businesses %to help their employees keep learning and developing themselves. A Udemy Business subscription 
that provides access to a curated subset of high-quality Udemy courses, via a separate Udemy Business API.% without having to pay for each course individually. 
\footnote{\url{https://business-support.udemy.com/hc/en-us/articles/11965611508375-Udemy-Business-API-Best-Practices}} %The Udemy Business API returns slightly different fields to the normal Udemy API. %The most relevant fields returned from the Udemy Business API are described in Table~\ref{tab:udemyb2b-api-fields}. 
%In total BrightFit has $10,414$ English courses from Udemy Business in its catalogue which are not already part of the standard Udemy set.
\textit{edX} is a company that offers high-quality university-level courses. %The courses on edX are created by some of the largest research institutions around the world like MIT and Harvard University. edX offers a selection of both free and paid courses. 
The data that we have available for the edX courses is limited compared to the data retrieved from the Udemy APIs. For the edX courses, we only have access to the course title, short course description and full course description. However, it is important to note that the short course descriptions of edX are of much better quality than the short course descriptions (headline) that we get from Udemy and Udemy Business. This makes them suitable for summarizer fine-tuning, as we will see later. %In total BrightFit has $2,077$ courses available from edX.
\textit{GoodHabitz} provides online courses and training as a subscription service for businesses. %The courses are created by the GoodHabitz team themselves which ensures that all the courses from them have a high-quality standard. 
The courses offered by GoodHabitz are largely focused on soft skills, like teamwork and collaboration. Similar to the edX courses the GoodHabitz courses only have a course title and course description to work with. %Compared to the other course providers for BrightFit, GoodHabitz is by far the smallest with only $127$ courses available.

\begin{table}[h]

\caption{Overview of the number of courses per provider.}
\label{tab:course-counts}
\begin{tabular}{l|r}
\hline
\textbf{Provider} & \textbf{Count} \\ \hline
Udemy             & 54,348          \\
Udemy Business    & 10,414          \\
edX               & 2,077           \\
GoodHabitz        & 127            \\ \hline
Total             & 66,966          \\ \hline
\end{tabular}
\end{table}

\subsection{Pre-processing}
\begin{figure}[t]
    \centering
    \includegraphics[width=.7\linewidth]{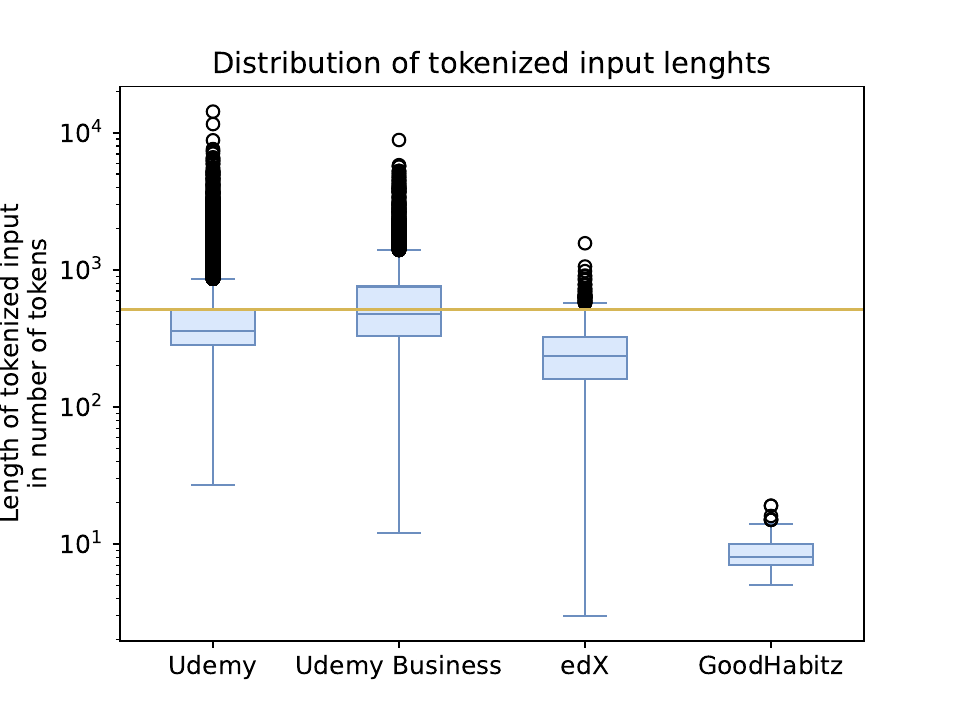}
    \caption{Overview of the tokenized input size distribution of the course titles combined with the descriptions of all BrightFit courses per provider. The horizontal line marks the input size of 512 tokens which is the maximum for T5.}
    \label{fig:input_sizes}
\end{figure}

We convert all course descriptions from HTML to plain text, harmonize the encoding to Unicode, and %To make it easier to work with the data collected from the different providers we apply some simple pre-processing. Many of the course descriptions contain HTML tags and other formatting characters. By applying a regex filter to the course descriptions we can quickly and effectively clean the data to simple plain text making it easier to work with. Some courses also have Unicode encoding issues which we also fix. Besides this, the only additional pre-processing that is applied is to 
standardize the names of the fields from the different course providers. %Table~\ref{tab:fields-overview} shows an overview of the standardized mapping we use to index all the courses from the different providers.
%Before we pass the input to the transformer-based models we need to tokenize the input. 
Figure~\ref{fig:input_sizes} shows a comparison of the lengths of the inputs for each course provider, consisting of the course titles concatenated with their full descriptions, tokenized with the Huggingface T5-base tokenizer~\cite{raffel-2019-t5}. We see that %the input lengths for many courses fall within the 512 length limit, but there are still many 
a substantial proportion (18,467 / 66,966 = 27.6\%) of courses are longer than %courses where the input lengths go beyond 
the limit of 512 tokens. By default, inputs that go beyond this limit will be truncated. This motivates the use of summarization (see Section~\ref{sec:summ})

% the input lengths of the courses. We can see that the majority of the courses fall within the 512 input token limit which is the limit for BERT and T5 models. There are however still a good number of outliers that go beyond this maximum input length.

\subsection{Queries} \label{sec:queries}
In BrightFit, users choose an occupation or job role they want to explore. They are asked to perform a self-assessment on how proficient they are in the most important skills for that role. After completing the self-assessment, users receive a report which includes the most important skills they need to develop, in order to have a more relevant skill set for the selected occupation or job role.
%Now that we know what the courses look like, we can focus on the queries we receive. 
For providing course recommendations, the BrightFit system %the queries are not directly issued by the users, but by the system based on the %skill gap that was found in the user assessment. 
generates a query based on the skill that the user wants to learn and their job/role title. The query that the retrieval engine receives has the form \textit{``<skill> for <occupation>''}, e.g. \textit{``Python for Data Analyst''}. Both terms come from a lookup in the Burning Glass taxonomy~\cite{Bonella_2020_BGOT} based on the self assessment of the user.\footnote{Burning Glass was later merged into Lightcast. We still use the original Burning Glass taxonomy.} %We can use this id to look up information about the skill like the name or description of the skill. 
%The second part of the query that we receive is the \textit{occupation id}, %. This occupation id corresponds to the new role that the user is currently exploring, and is 
%also taken from the Burning Glass taxonomy. 
The occupation is used to provide more context for the course recommendations. %This is useful when we want to make recommendations for a skill that can be applied in different occupations, like the Python programming language. If we have knowledge about the occupation, role, or job that the user is exploring, we can make better recommendations for Python courses that are more relevant. For example, it is a lot more useful to recommend a Python course for data analysis if the user is exploring a data analyst role, compared to when they would be exploring the role of a backend developer.
%An example of a structured query is \textit{\{"skill\_id": "4202", "occupation\_id": "US1234"\}}. A lookup in the skills taxonomy provides the skill \textit{"Python"} and the occupation \textit{"Data Analyst"}. %We can use these ids to look up information in the taxonomy like the English labels that correspond to the skill and occupation like we just did for \textit{Python} and \textit{Data Analyst}. We can also look up more information like the skill description for the \textit{Python} skill which is as follows: \textit{"Python is a widely used high-level programming language for general-purpose programming, created by Guido van Rossum and first released in 1991."}

\subsection{Evaluation data}
In this section, we %discuss the datasets that we use for the evaluation of our methods. We 
present two datasets that we created for evaluating BrightFit: the BrightFit IT dataset and the BrightFit general skills dataset.

\subsubsection{BrightFit IT dataset}
%To evaluate how well our methods work on BrightFit data, we create our own datasets based on BrightFit data. 
The first dataset that we create is the BrightFit IT dataset, which is only focussed on IT skills. IT skills often have more courses available compared to other types of skills. This should give us a good dataset that can be used to see how the system behaves in cases where there are plenty of courses available for a skill.

%We first discuss how we create the queries of the dataset. Next, we discuss how we obtain the list of documents for each query for which we produce a relevance label. Lastly, we look at how we assign a relevance label to each query-document pair.

\paragraph{Query generation}
The first step in creating the dataset is to select the queries we will use to evaluate the model. %For the queries, we focus on skills and occupations related to IT. By using IT occupations and skills it is easier for us to give better relevance judgements as we have more domain knowledge about the IT sector compared to other sectors. In addition to this, IT skills are often easier to learn online compared to soft skills or more practical skills which can be more common in other sectors. 
The course catalogue of BrightFit has many courses related to IT skills (because IT skills are relatively easy to learn online).
In total we label data for 15 different skills. %, but we also want to focus on the occupation aspect of the course recommendations. Therefore 
We combine each skill with 3 occupations that are common for that skill giving us a total of $15 * 3 = 45$ queries for our dataset.
%The first step is to get the IT occupations from the taxonomy. This is easy to do since IT occupations have their own subgroup in the Burning Glass Occupation Taxonomy. 
We select the IT occupations from the Burning Glass Occupation Taxonomy. % We apply some additional filtering to the occupations obtained to get a good variety of occupations: 
We improve the diversity of our set by filtering out the senior versions of other occupations that are in the taxonomy, e.g. ``Senior Java Developer/Engineer" when ``Java Developer/Engineer" also exists,  %in terms of recommendations and we are better off looking at more different occupations to evaluate the models on more varying data. 
%In total, we have 153 IT occupations and after filtering out the Senior versions of titles we are left with 149 IT occupations. 
and we filter out the very general occupation ``Software Developer/Engineer". %, because after initial testing we observed that this occupation dominated all the others in terms of occurrences for skills; We conclude that this occupation is used as a general term for a variety of software developer jobs such as "Java Developer" ans "Python Developer". %For this reason, 
%We chose to %leave out the general "Software Engineer/Developer" and instead 
%prioritize the more specific developer occupations. % since they are also more specific and make it easier to judge the relevance of a course for an occupation, leaving us with 
The resulting set contains 148 IT occupations.
%Once we have the filtered list of IT occupations, we 
We then select for each of the 15 skills the 3 occupations that have the most mentions for the skill in the Burning Glass job posting data about skill occurrences in the United States~\cite{lightcast-data} in 2022. %\footnote{The skill set in the Lightcast API is different than the original Burning Glass skills. The reference to the BG API is no longer available due to their merger. The type of data available from the Lightcast API is very similar to the original Burning Glass data.} 
This way, we create 45 skill--occupation pairs.

\paragraph{Document pool creation}
We manually label 50 documents for each query. %We chose this number because we believe that this gives a nice balance between still having enough documents to re-rank the order, but also still having a manageable workload to label by 1 person. 
Since one of the use cases for the annotated dataset is to compare the current BrightFit retriever with the newly proposed method, we need to create a fair dataset for both rankers. To accomplish this we obtain the top 25 from the two retrievers we discuss in Section~\ref{sec:1ststage}. We interleave documents retrieved from the two retrievers to build a top 50 list. We remove duplicate documents for queries where both retrievers have retrieved the same document. %By interleaving the documents retrieved from both approaches we keep the top 50 list for each query as unbiased as possible.

\paragraph{Relevance assessment}
%The final step in creating the dataset is to assign relevance labels to the query-document pairs that we created in the previous steps. For the labels we have decided to use 3 different relevance levels:
We label the query--document pairs on three relevance levels:
\begin{itemize}[leftmargin=*]
    \item \textbf{2}: Course is relevant for the skill--occupation combination and specific to the occupation. E.g. with the query ``Python for ML Engineer" and a course ``Learn Python for Machine Learning".
    \item \textbf{1}: Course is relevant for the skill but it is generic. E.g. a course ``Python for beginners" for the query ``Python for ML Engineer".
    \item \textbf{0}: Course is not relevant for the skill or the course is specific for the skill and a different occupation. E.g. When the query is ``Python for Database Engineer" and the course is "Learn Python for Machine Learning".
\end{itemize}

The idea behind the labels is as follows: In the ideal case we want to recommend courses to users that teach a specific skill, but that are also relevant to the occupation they are trying to close their skill gap to. Therefore we want to prioritize courses that teach a skill in a context that is relevant for the user. However, it will often be the case that there is no perfect course that exactly teaches a certain skill in the context of the specific occupation we are looking for. In this case, we need alternatives to recommend that still teach the skill. %, as we need to make distinctions between the "perfect" courses, good alternatives that teach the skill in a generic manner, and irrelevant courses. Lastly, we have the irrelevant courses. 
We assign label 0 to courses that do not teach the skill we are searching for and %This is rather straightforward as the content of the course does not match our query. Besides assigning label 0 to courses that are not about the skill we are looking for, we also assign label 0 to 
courses that do teach the skill but in a context that is very different from the role the user is exploring. We do this as we believe that a generic course about a skill is more relevant than a course that teaches a skill in a context that is not relevant to the user. These labelling guidelines have also been discussed with a Learning and Development expert affiliated to BrightFit, % from Randstad, 
to ensure that the guidelines align with what is expected from the recommender system.

%Appendix~\ref{app:labelling-examples} contains a handful of examples of queries and documents with their assigned label and an explanation of how we arrived at the labels. 

\subsubsection{BrightFit general skill dataset}
%To also get an idea of how the changes to the recommender system affect the general course retrieval in BrightFit and not only specific IT-related retrieval, we also create a smaller dataset with random queries from all possible skills and occupations.

Similar to the BrightFit IT dataset we first create a set of queries. For this dataset, we randomly sample 10 occupations from the full taxonomy. We then select a random skill from the top 10 most common skills for each occupation based on the Burning Glass job posting data~\cite{lightcast-data}. %After some initial observations of the query--document pairs produced by this method, we have 
We added as an additional constraint to the random skill/occupation pairs that the first-stage retriever should be able to at least retrieve 5 documents with a cosine similarity >= 0.6. This additional constraint improves the quality of the queries and documents in the dataset, making the dataset more useful for evaluation. To obtain the documents to rate for each query we use the same method as for the BrightFit IT dataset. We also use the same relevance labelling rules.

Table~\ref{tab:evaldata} shows the statistics for the two datasets. On average, the BrightFit IT skills dataset has more relevant courses for each query than the general skills dataset. This is because there exists a large amount of online courses for IT-related skills as they are easier to teach and learn online. %However, we still see that there are plenty of relevant courses available for the queries in this general skills dataset. 

\begin{table}[h]
\centering
\caption{Overview of the two labelled BrightFit datasets. The total number of indexed documents in BrightFit is 66,966.}
\label{tab:evaldata}
\begin{tabular}{lrr}
\hline
Dataset information & IT & general \\ \hline
%Total number of documents in collection & 66,966 & 66,966 \\
Total number of unique documents with label & 1,035 & 500 \\
Total number of queries & 45 & 10 \\ \hline
Query information &  & \\ \hline
Lowest average relevance per query & 0.20 & 0.26\\
Highest average relevance per query & 1.68 & 1.20\\
Mean average relevance per query & 0.84 & 0.74\\
Mean relevance count per query, label 0 & 19.3 & 24.0 \\
Mean relevance count per query, label 1 & 19.3 & 14.9\\
Mean relevance count per query, label 2 & 11.4 & 11.1\\
Average number of labels per query & 50 & 50  \\ \hline
\end{tabular}
\end{table}

\section{Methods}\label{sec:methods}

We follow a two-stage retrieval approach, with efficient first-stage retrievers (\ref{sec:1ststage}), and a more effective re-ranker (\ref{sec:rerank}). We experiment with summarization of course descriptions (\ref{sec:summ}) and quantization of the re-ranker for more efficient real-time ranking (\ref{sec:quant}).

\subsection{First-stage retrieval} \label{sec:1ststage}

The current in-production ranker in the BrightFit course recommendation system is BM25 with a weight of 2 for the title and 1 for the course description. Additionally, a score multiplier of 7 is applied when there is a perfect match with the skill text in the title. We compare this in-production ranker to a T5-based~\cite{raffel-2019-t5} retriever called GTR~\cite{ni2021gtr} which has shown good zero-shot performance on the BEIR benchmark. In this research, we use the \texttt{GTR-Base}\footnote{\url{https://huggingface.co/sentence-transformers/gtr-t5-base}} version of the retriever as a zero-shot first-stage retriever. The query we use for first-stage retrieval is the same query format as described in Section~\ref{sec:queries}: ``\textit{<skill> for <occupation>}''. The documents are embedded using the format: \textit{``Title: <course title> Description: <course description>"}. For a given query we retrieve the top $k$ courses based on cosine similarity.

\subsection{Re-ranking} \label{sec:rerank}
As our second stage re-ranker model, we use RankT5~\cite{zhuang2022rankt5}. We choose the RankT5-Enc architecture since this has half the number of parameters compared to the RankT5-EncDec architecture while achieving comparable performance~\cite{zhuang2022rankt5}. With fewer parameters, the model also uses less memory and has faster inference speed which is useful for the efficient deployment of the model in the BrightFit application.

We do not have access to labelled data that we can use for fine-tuning the RankT5 model to the BrightFit domain, but it was shown~\cite{zhuang2022rankt5} that the model is capable of achieving good zero-shot performance by training it on the MSMARCO passage ranking dataset~\cite{bajaj2018msarco}. We use the version of the MSMARCO dataset in the BEIR benchmark~\cite{thakur2021beir} to finetune our model. % and then we apply inference on the BrightFit data without any further training on BrightFit-specific data.
Finetuning RankT5 requires hard negatives. We sample these hard negatives for each query in the MSMARCO dataset using the GTR-Base retriever to obtain the top 1000 documents for each query and consider each document that does not have a positive relevance label for the query as a hard negative. 

We use the listwise softmax loss function %it has better zero-shot performance than the pointwise loss that is also discussed in the RankT5 paper. We use 
with a list size of 36 for the softmax loss, as higher list sizes have shown to result in better zero-shot performance~\cite{zhuang2022rankt5}. We train the model for 50,000 steps, similar to the RankT5 paper, and we evaluate the model every 5,000 steps and take the best-performing checkpoint. For training, we use the AdamW optimizer with a constant learning rate of $10^{-4}$. Due to hardware limitations, we are not able to run a batch size of 32 like the RankT5 authors, but instead, we run a batch size of 2 with 16 steps of gradient accumulation to emulate the batch size of 32. We implement our model in the PyTorch framework and use the T5-base checkpoint from HuggingFace\footnote{\url{https://huggingface.co/t5-base}} to initialize our encoder. We fine-tune our RankT5 model using an RTX 3090 GPU which takes up to 100 hours to fully train the model for 50,000 steps.

The input format for the re-ranker combines the query and document format that we use for the first-stage retriever and combines them using the format described by the RankT5 authors: \textit{``Query: <skill> for <occupation> Document: Title: <course title> Description: <course description>''}. If the input to the re-ranker is longer than the maximum input size (512 tokens by default) the input is truncated.

\subsection{Summarization} \label{sec:summ}
Since a substantial proportion of the course descriptions in our collection are too long to fit the 512-token input length that T5 supports (see Figure~\ref{fig:input_sizes}), we investigate automatic summarization of the course descriptions as an alternative to truncating. %, to potentially improve the effectiveness of the retrieval. 
%As a way to improve the performance of the second-stage re-ranker, we evaluate the use of summarization of the course descriptions. We perform the re-ranking of the courses using these summarized descriptions instead of the original course descriptions. By applying summarization to the course descriptions we aim to achieve two goals. The first goal is to reduce the length of descriptions that are too long to fit into the 512 token-long input length that T5 supports, in a more informative way than by using truncating. %When documents are longer than the 512 maximum input length the input will get truncated. This can cause us to throw away valuable information about the course content that would help with re-ranking the document. By applying summarization, we aim to capture the essential parts of the course contents in a few short sentences, while leaving out the parts that do not contribute to judging the relevance of a course. 
We hope that a side effect of summarization is that we dispose of irrelevant information in the course descriptions that provides noise to the ranker. Examples of this noise are positive user reviews included by course instructor to promote their course on the platform, %An example of such a description with user reviews is shown in Appendix~\ref{app:course_desc}. Next to reviews in the course description, there are also many other examples of useless information in the descriptions, like 
information about which well-known companies are using the course, and information about the Udemy 30-day money guarantee. Although this information can be relevant to users, it is noise to the ranker. By using summarization we aim to give the re-ranker a cleaner input, which will hopefully improve the ranking performance. The users will be presented with the full course descriptions in BrightFit.

We generate the summaries at index time, together with the embeddings used in first-stage retrieval. This means that we do not add any additional computation at query-time. 

Transformer-based sequence-to-sequence models like T5 have shown strong performance on summarization tasks~\cite{raffel-2019-t5} and %But they are limited to an input length of only 512 tokens. 
summarization techniques based on the Longformer encoder-decoder model (LED)~\cite{beltagy2020longformer} have effectively been applied to long documents in the context of case law retrieval~\cite{althammer2021dossiercoliee, Askari2021CombiningLA}. 

We experiment with two models for summarization: LongT5~\cite{guo2022longt5} and Vicuna~\cite{vicuna2023}. \textbf{LongT5} is a T5 model that has been adapted to work efficiently for longer input sequences. %The authors accomplished this by introducing a new attention mechanism called \texttt{TGlobal} which is a method for combining local and global attention in T5 and other transformer models. 
The model has shown good performance on summarization of long documents compared to standard T5 models. We use the \texttt{long-t5-tglobal-base}\footnote{\url{https://huggingface.co/google/long-t5-tglobal-base}} checkpoint from HuggingFace as the base model. We fine-tune this model on 1,868 pairs of long and short course descriptions from the edX data. We chose the edX course descriptions since the short course descriptions from edX are of higher quality than the short descriptions from Udemy. %\footnote{During our analysis, we found that mainly the courses from Udemy were very noisy and would likely benefit the most from the summarization techniques. However, we did not have proper training data available to fine-tune the LongT5 model for specifically summarizing the Udemy course descriptions. Therefore, we decided to use the edX descriptions for fine-tuning the summarizer.} 
We exclude from the training data the edX course descriptions that are part of our labelled test collection, to prevent data leakage from the train to the test set. 
We fine-tune the model for 2 epochs on the pairs of full and short descriptions. We use a learning rate of $10^{-4}$ and a batch size of 1 with 16 gradient accumulation steps. We select the model with the highest \textit{ROUGE} score which we evaluate at the end of each epoch. We limit the maximum input length to 2048 to keep it comparable with the Vicuna summaries.

The \textbf{Vicuna} models are a collection of LLMs that are adaptations of Llama~\cite{touvron2023llama}, %The Vicuna models were trained by
fine-tuned on ChatGPT conversations shared by users on the \url{ShareGPT.com} website. %Another important change to Vicuna compared to the Llama models is the increased context length. 
The Vicuna maximum input context length is 2048 tokens, which is useful for our summarization use-case. The Vicuna model that we use is one of the Vicuna 7B checkpoints\footnote{\url{https://huggingface.co/lmsys/vicuna-7b-v1.3}} based on the 7B parameter Llama model. The main reason for selecting the 7B model is that it is still possible to fit this model in 8-bit format on a single 16GB GPU and perform inference with it. We use the Vicuna model as a zero-shot summarizer by providing a fixed input prompt along with a course title and a course description for generating the summary. %Based on earlier observations on the course description and some small experiments, we formulated the 
The prompt that we used is shown in Figure~\ref{fig:prompt}. When the input is longer than the 2048 input window we truncate the input. We limit the number of generated tokens to 256 so that we are sure we are getting shorter course descriptions that easily fit into the input window of our re-ranker.

\begin{figure}[t]
{\small
\begin{tabular}{p{15cm}}
\texttt{I will provide you with a course title and description of an online course that I want you to summarize in 2 to 3 lines.}\\
\texttt{I want the summary to only include information about the content of the course. 
You can leave out any information about the author, at which company the course is used and information about a 30 day money back guarantee.
You can also leave out any student reviews about the course. I want you to write the summary as if it were a new shortened course description.}\\
\texttt{Course title: [course\_title]}\\
\texttt{Course description: [course\_description]}\\
\end{tabular}}
\caption{Prompt used for Vicuna to summarize course descriptions}\label{fig:prompt}
\end{figure}

\subsection{Quantization} \label{sec:quant}
The main bottleneck for inference speed in our approach is the re-ranker. To reduce this bottleneck, we experiment with 3 different quantization methods: dynamic quantization, static quantization~\cite{Gholami2021quantizationsurvey}, and SmoothQuant~\cite{xiao2023smoothquant}. %We choose to use SmoothQuant over other more advanced and better-performing quantization methods such as AWQ~\cite{lin2023awq} to avoid hardware support issues that arise when working with low-bit quantization methods like AWQ. 
We use the implementation of SmoothQuant in the Neural Compressor package by Intel.\footnote{\url{https://github.com/intel/neural-compressor}} %The Neural Compressor package has support for many different quantization methods and deep learning frameworks. During our initial experimentation with these quantization methods in the Neural Compressor package, we found that not all quantization methods are completely stable for the PyTorch framework as the package is under constant development. During some initial experiments, we found that the implementations for static quantization, dynamic quantization and SmoothQuant were the most stable with the ONNX framework. We therefore export our trained model to the ONNX format before we apply quantization using the Neural Compressor package.
To perform the static quantization and quantization using the SmoothQuant algorithm we need to have a calibration dataset. To stay true to the zero-shot setting we use the \textit{dev} partition of the MSMARCO dataset to randomly select samples for calibration instead of using BrightFit data.
We use the version of SmoothQuant %the authors proposed 3 different variations. In this research, we use the version 
with the per-tensor dynamic quantization for the activations because the alternative versions have slower inference speed or lead to lower quality output according to the SmoothQuant paper~\cite{xiao2023smoothquant}. %We use this version since both alternatives have some drawbacks. One of the alternatives is per-token dynamic activation quantization. This alternative has comparable performance to the per-tensor dynamic quantization but the inference speed is slower. The other alternative is per-tensor static quantization. This alternative is faster than the version we choose but the authors see a quality degradation when using this method. 

\section{Results}
\label{sec:experiments-and-results}

We use the BM25-based in-production ranker in the BrightFit course recommendation system as a baseline in our ranking evaluation. 

\subsection{Experiment 1: First-stage retrieval}
%In this experiment, we compare the two first-stage retrievers: BM25 and GTR. %With this experiment, we also aim to answer \textbf{[RQ1]}: \textit{"How suitable are transformer-based first-stage retrievers for our use case?"}
%We evaluate the retrievers on both of the labelled BrightFit datasets. %We do not evaluate the BM25 and GTR methods on the BEIR benchmark as we use both algorithms with minor modifications and they have both been compared to each other on the BEIR benchmark before in~\cite{ni2021gtr}.
Table~\ref{tab:res-first-stage} shows the results of the in-production BM25 baseline and the GTR first-stage retriever on the two BrightFit datasets. The table shows that GTR outperforms BM25 by a large margin. For the first-stage retrievers, the metric we are most interested in is recall, since we want the candidate list to have as many relevant documents as possible to obtain a better final ranking after re-ranking the candidate list using the second-stage re-ranker. %We see that also for this metric the GTR retriever is performing better than the BM25 retriever. 
%In the first-stage retrieval, recall is an important metric, because the re-ranking stage will not retrieve any new documents. 
The results for different levels of recall are shown in Table~\ref{tab:res-exp4-bf-recall}. The table shows that the GTR retriever is consistently better than the BM25 retriever for all tested values of $k$. %We also observe that there are significant gaps in terms of recall score between the different values of k, showing us how increasing the value of k can give us a lot more relevant documents in our candidate set. For this part of the experiment we only have rated up to the top 25 for each retriever. Therefore we assume that all the documents that we retrieve that are not rated have a relevance of 0.

\begin{table}[t]
\centering
\caption{Comparison between the BM25 and GTR-Base retriever on the two BrightFit datasets. BM25 is the current in-production ranker.}
\label{tab:res-first-stage}
\centerline{
{\small
\begin{tabular}{lccccc}
\hline
Ranker & Dataset & nDCG@10 & MRR@10 & MAP@10 & R@20 \\ \hline
BM25 & IT & 0.482 & \textbf{0.931} & 0.139 & 0.380 \\
 & General & 0.447 & 0.778 & 0.150 & 0.400 \\ \hline
GTR & IT & \textbf{0.648} & \textbf{0.931} & \textbf{0.221} & \textbf{0.512} \\
 & General & \textbf{0.705} & \textbf{1.000} & \textbf{0.311} & \textbf{0.545} \\ \hline
\end{tabular}}
}
\end{table}

\begin{table}[t]
\centering
\caption{Recall@K scores for the BM25 and GTR retrievers on both BrightFit datasets for different values of k.}
\label{tab:res-exp4-bf-recall}
\begin{tabular}{lllll}
\hline
Dataset & Ranker & k=20  & k=30  & k=50  \\ \hline
IT      & BM25      & 0.380 & 0.504 & 0.555 \\
        & GTR       & \textbf{0.512} & \textbf{0.661} & \textbf{0.739} \\ \hline
General & BM25      & 0.400 & 0.498 & 0.548 \\
        & GTR       & \textbf{0.545} & \textbf{0.651} & \textbf{0.668} \\ \hline
\end{tabular}
\end{table}

\subsection{Experiment 2: Re-ranking}\label{sec:reranking}
%In this second experiment, we evaluate our second-stage re-ranker. %We also aim to answer our second underlying research question \textbf{[RQ2]}: \textit{"How effective are zero-shot re-ranking methods for ranking course recommendations?"}
We evaluate RankT5 with different configurations to see what variation gives us the best performance on both the BrightFit IT dataset and the BrightFit general skills dataset. The first aspect of the model that we experiment with is the maximum input length. We experiment with three different values: 128, 256 and 512 tokens. The input that goes past this maximum input length is truncated. The second aspect we vary is whether we do or do not include the short skill description from the Burning Glass taxonomy in the query for re-ranking. By adding the skill description to the query we can add more context, but it can also confuse the model since the input queries become much longer. This also means that we have less space for the document text in the limited input window of the encoder which can also hurt performance. 

During the training process, we observe that the RankT5 model obtains its best zero-shot performance at 25k training steps, so for this experiment and the following, we use that checkpoint. The results of this experiment are shown in Table~\ref{tab:res-exp-2}. %We re-rank the top 50 documents that we obtained and labelled using the interleaved list of the 2 retrievers. 
We observe that for both datasets the best-performing settings are the ones where we have a maximum input length of 128 and we do not include the skill descriptions. There is a substantial performance drop for all model lengths if we include the skill description in the query.

\begin{table}[t]
%\vspace*{-12pt}
\centering
\caption{nDCG@10 scores of the second-stage re-ranker on both BrightFit datasets for different variations compared to the in-production BM25 ranker.}
\label{tab:res-exp-2}
\centerline{
{\small
\begin{tabular}{l|lll|lll}
\hline
Dataset                           & \multicolumn{3}{c|}{BrightFit IT} & \multicolumn{3}{c}{BrightFit General} \\ \hline
Max input length & \multicolumn{1}{c}{128} & \multicolumn{1}{c}{256} & 512 & \multicolumn{1}{c}{128} & \multicolumn{1}{c}{256} & \multicolumn{1}{c}{512} \\ \hline
W/ skill description  & 0.613           & 0.628  & 0.612  & 0.627             & 0.696   & 0.646   \\
W/o skill description & \textbf{0.682}  & 0.660  & 0.626  & \textbf{0.798}    & 0.758   & 0.729   \\ \hline
\textit{BM25 (in-production)}                           & \multicolumn{3}{c|}{\textit{0.482}} & \multicolumn{3}{c}{\textit{0.447}} \\ \hline
\end{tabular}}}
\end{table}

We also analyzed the impact of changing the re-ranking depth on the final ranking performance. %We are limited to a maximum re-ranking depth of 50, as this is the number of courses we have rated for each query. 
We experiment with the following re-ranking depths: 50, 30 and 20, and compare the NDCG@10 scores after re-ranking for both BrightFit datasets. %To evaluate the re-ranking we use the interleaved candidate set of both retrievers that we labelled instead of using the top-k from either the GTR retriever or the BM25 retriever because we only have relevance labels up to the top 25 for each of the retrievers. By using the interleaved list we have a true relevance label for each of the 50 documents we re-rank. 
The results are shown in Table~\ref{tab:res-exp4-bf-ndcg}. For both datasets, we obtained the highest NDCG@10 score with a re-ranking depth of 30. %We notice that the difference in NDCG@10 scores between different values of k is especially large for the general skills dataset, but there is also a substantial difference for the IT dataset. %This shows us that the re-ranking depth is an important parameter to get right.

\begin{table}[t]
\centering
\caption{NDCG@10 scores after re-ranking the top-k documents with RankT5 retrieved on both BrightFit datasets for different values of k.}
\label{tab:res-exp4-bf-ndcg}
\begin{tabular}{llll}
\hline
Dataset & k=20 & k=30 & k=50 \\ \hline
IT & 0.675 & \textbf{0.702} & 0.677 \\
General & 0.778 & \textbf{0.881} & 0.844 \\ \hline
\end{tabular}
\end{table}

\subsection{Experiment 3: Summarization}
We evaluate the effectiveness of ranking summarized versions of the course descriptions. We compare three settings: the default course description (truncated if needed), a summary generated by our fine-tuned LongT5 summarizer, and a summary generated by the zero-shot Vicuna model. We also explore how the effectiveness of the ranker with summarized descriptions changes when we vary the maximum input length of the ranker. We experiment with the values of 128, 256, and 512 for the maximum input length. %We evaluate the summarization methods on both the BrightFit datasets. %With this experiment we also aim to answer our next underlying research question \textbf{[RQ3]}: \textit{"Can we improve ranking performance by applying summarization techniques to the course descriptions?"}
The results of this experiment are shown in Table~\ref{tab:res-exp-3}.

\begin{table}[t]
\centering
\caption{NDCG@10 scores of the RankT5 re-ranker on both BrightFit datasets for different summarization techniques.}
\label{tab:res-exp-3}
{\small
\begin{tabular}{l|lll|lll}
\hline
Dataset        & \multicolumn{3}{c|}{BrightFit IT} & \multicolumn{3}{c}{BrightFit General}   \\ \hline
Max length & \multicolumn{1}{c}{128} & \multicolumn{1}{c}{256} & 512 & \multicolumn{1}{c}{128} & \multicolumn{1}{c}{256} & \multicolumn{1}{c}{512} \\ \hline
Original    & 0.682           & 0.660  & 0.626  & 0.798 & 0.758          & 0.729          \\
LongT5  & 0.680 & \textbf{0.684}  & 0.683  & 0.814 & 0.801          & 0.798          \\
Vicuna  & 0.669           & 0.676  & 0.677  & 0.806 & \textbf{0.844} & \textbf{0.844} \\ \hline
\end{tabular}}
\end{table}

We find that on both BrightFit datasets the NDCG@10 score improves by summarizing the course descriptions. For the BrightFit IT dataset, we obtain the best results using the LongT5 summarizer for an input length of 256. Overall, we see that the improvement on the BrightFit IT dataset is very minimal compared to the default description with an input length of 128. Interestingly, we also observe that the Vicuna summaries on the BrightFit IT dataset do not directly outperform the 128 input length for the standard description. We do, however, notice that when the input lengths get longer the performance of the standard description starts to decrease, while the performance for the Vicuna summaries starts to increase. We think this is because Vicuna summaries are created to contain the essence of the course, leaving out irrelevant information; the ranker therefore benefits from including a longer summary. The original descriptions on the other hand are noisy and contain irrelevant information later in the text, so adding more of the description adds more irrelevant information.

%For the BrightFit general skills dataset, we find that the Vicuna-generated summaries achieve the best performance at the input lengths of 256 and 512. 
We also see that the LongT5 summaries make an improvement over the standard description for this dataset but not as much as Vicuna. The improvement we get on the general skills dataset is much larger than the improvement we get on the IT dataset. %We also see that the gap between LongT5 and Vicuna is much larger on the BrightFit general skills dataset than the difference on the BrightFit IT dataset.
% TODO mention improvement over description
% TOD mention that vicuna is not that far off
% For discussion: hard skills are easier to include in the summary?

\subsection{Experiment 4: Quantization} \label{sec:expquant}
%We evaluate the impact of quantization on our models. 
We evaluate the static, dynamic and SmoothQuant quantization methods on model size and inference speed. SmoothQuant has a single parameter $\alpha$, which is the smoothing factor. We experiment with three values of $\alpha \in \{0.25,0.50,0.75\}$, with 0.5 being recommended by the authors for most models. Since the results for the three values are very close to each other, we only report the results for $\alpha=0.5$. The static and dynamic quantization algorithms do not have any parameters.

We look at how much space a model takes up after quantization compared to the full FP32 model and we look at how the inference speed is improved by measuring the throughput of the model. %With this experiment, we aim to answer our fifth underlying research question \textbf{[RQ4]}: \textit{"What effect do different quantization methods have on the inference speed and prediction quality?"}
%For quantizing the models, we use the Intel Neural Compressor package. As a calibration dataset for static quantization and SmoothQuant, we use random samples from the MSMARCO dataset dev partition. For SmoothQuant we use $\alpha=0.5$ based on the recommendations made by the authors of the SmoothQuant paper~\cite{xiao2023smoothquant}. 
We use the built-in benchmarking tool from the Intel Neural Compressor package for this. We benchmark the models using 2 cores per model. For each quantization method, we run 8 repetitions. We use a batch size of 16. In each repetition, we use a warmup of 10 inference queries after which we run 200 inference query--document pairs from the BrightFit general skills dataset and measure the throughput of the model as the number of query--document pairs that are processed per second. Since measuring inference speed is very hardware- and sometimes even package-dependant we report all the used hardware and package versions in Table~\ref{tab:versions}.

\begin{table}[h]
\centering
\caption{Hardware and package version overview for the quantization experiments.}
\label{tab:versions}
\begin{tabular}{ll}
\hline
Hardware                & Version \\ \hline
Google Cloud Platform VM & \begin{tabular}[c]{@{}l@{}}n1-standard-16 VM \\ (16 vCpus, 60 GB RAM)\end{tabular} \\ \hline
Packages                & Version \\ \hline
intel-neural-compressor & 2.3     \\
onnxruntime             & 1.16.0  \\
onnx                    & 1.14.1  \\
transformers            & 4.30.2  \\ \hline
\end{tabular}
\end{table}

Table~\ref{tab:res-exp5-quant-speed} shows the inference speed and model sizes of the different quantization methods and the non-quantized FP32 model. %We record the model sizes for all quantization methods and we measure the throughput in queries per second for different quantization methods and input sizes. 
We see that the dynamic quantization method obtains the best throughput for both the 256 and 512 input lengths. We observe over 40\% speed-up of the dynamic quantization method compared to the default FP32 model. In terms of throughput the static and SmoothQuant quantized models are not far behind the dynamically quantized model. For the dynamic and SmoothQuant methods, we also observe a close to 4x reduction in model size which is as expected since we convert the model weights from 32-bit to 8-bit values. %We see that the static quantization method takes up slightly more space than the other quantized models which is the result of the static quantization method having to store the scaling factors and zero-point offsets in the quantized model.

% \begin{table}[t]
% \centering
% \caption{Throughput in queries per second for different input length sizes (256 and 512) and model size (in MB) comparisons for different quantization methods. The standard deviations are shown in brackets.}
% \label{tab:res-exp5-quant-speed}
% \begin{tabular}{lccc}
% \toprule
% Quantization & Throughput (256) & Throughput (512) & Model size \\
% \midrule
% None (FP32) & 2.293 (\textpm 0.040) & 1.190 (\textpm0.002) & 418  \\
% Static & 3.155 (\textpm0.054) & 1.631 (\textpm0.003) & 128.8 \\
% Dynamic & \textbf{3.381 (\textpm0.016)} & \textbf{1.696 (\textpm0.015)} & \textbf{105.2} \\
% SmoothQuant & 3.259 (\textpm0.012) & 1.549(\textpm0.011) & 105.5 \\
% \bottomrule
% \end{tabular}
% \end{table}

\begin{table}[t]
\centering
\caption{Throughput in queries per second for different input length sizes (256 and 512) and model size (in MB) comparisons for different quantization methods. The standard deviations are shown in brackets.}
\label{tab:res-exp5-quant-speed}
\begin{tabular}{lccc}
\hline
Quantization           & \multicolumn{1}{l}{Throughput (256)} & \multicolumn{1}{l}{Throughput (512)} & \multicolumn{1}{l}{Model size} \\ \hline
None (FP32)            & 2.573 ($\pm 0.035$)                     & 1.226 ($\pm 0.034$)                     & 418.4                          \\
\textit{Static}        & 2.875 ($\pm 0.079$)                     & 1.545 ($\pm 0.043$)                     & 128.8                          \\
Dynamic                & \textbf{3.420} ($\pm 0.046$)                     & \textbf{1.582} ($\pm 0.029$)                     & \textbf{105.2}                          \\
%SQ ($\alpha=0.25$) & 3.180 ($\pm 0.034$)                     & 1.513 ($\pm 0.026$)                     & 105.5                          \\
SmoothQuant & 3.292 ($\pm 0.050$)                     & 1.528 ($\pm 0.023$)                     & 105.5                          \\
%SQ ($\alpha=0.75$) & 2.843 ($\pm 0.178$)                     & 1.560 ($\pm 0.013$)                    & 105.5                          \\ 
\hline
\end{tabular}
\end{table}

% \begin{table}[t]
% \centering
% \caption{NDCG@10 scores of various quantization methods on the BrightFit datasets compared to the standard, non-quantized FP32 model as a reference.}
% \label{tab:res-exp5-quant-scores}
% \begin{tabular}{l|cc|cc}
% \hline
% Dataset          & \multicolumn{2}{c|}{BrightFit IT} & \multicolumn{2}{c}{BrightFit General} \\ \hline
% Max input length & 256             & 512             & 256               & 512               \\ \hline
% \textit{RankT5--FP32}             & \textit{0.659}            & \textit{0.623}           & \textit{0.754}             & 0.\textit{725}    \\
% RankT5--Static           & 0.503           & 0.511           & 0.444             & 0.389             \\
% RankT5--Dynamic          & \textbf{0.657}           & \textbf{0.636}  & \textbf{0.775}    & \textbf{0.720}             \\
% RankT5--SmoothQuant      & 0.624           & 0.595           & 0.732             & 0.692             \\ \hline
% \end{tabular}
% \end{table}
\begin{table}[t]
\centering
\caption{NDCG@10 scores of various quantization methods on the BrightFit datasets compared to the standard, non-quantized FP32 model as a reference. * indicates that the difference with the FP32 model is statistically significant according to a paired t-test with $\alpha=0.05$ and Bonferroni correction for multiple testing.}
\label{tab:res-exp5-quant-scores}
\begin{tabular}{l|cc|cc}
\hline
Dataset                       & \multicolumn{2}{l|}{BrightFit IT} & \multicolumn{2}{l}{BrightFit General} \\ \hline
% Max input length              & 256            & 512           & 256             & 512              \\ \hline
% \textit{RankT5-FP32}          & \textit{0.659} & 0.623         & 0.754           & \textit{0.725}   \\
% RankT5-Static                 & 0.503 (.0002)  & 0.511 (.0106) & 0.444 (.0014)   & 0.389 (.0075)    \\
% RankT5-Dynamic                & 0.657 (.7763)  & 0.636 (.0444) & 0.775 (.2918)   & 0.720 (.6139)    \\
% RankT5-SQ ($\alpha=0.25$) & 0.626 (.0756)  & 0.596 (.1462) & 0.732 (.3960)   & 0.682 (.0542)    \\
% RankT5-SQ ($\alpha=0.50$) & 0.626 (.0740)  & 0.597 (.1503) & 0.737 (.4690)   & 0.691 (.2038)    \\
% RankT5-SQ ($\alpha=0.75$) & 0.624 (.0569)  & 0.604 (.2554) & 0.726 (.2460)   & 0.692 (.1890)    \\ \hline
Max input length              & 256            & 512           & 256             & 512              \\ \hline
\textit{RankT5-FP32}          & \textbf{0.659} & 0.623         & 0.754           & 0.725   \\
RankT5-Static                 & 0.503*   & 0.511 & 0.444*   & 0.389    \\
RankT5-Dynamic                & 0.657   & 0.636 & \textbf{0.775 }  & 0.720    \\
%RankT5-SQ ($\alpha=0.25$) & 0.626   & 0.596 & 0.732   & 0.682    \\
RankT5-SmoothQuant & 0.626  & 0.597 & 0.737   & 0.691    \\
%RankT5-SQ ($\alpha=0.75$) & 0.624  & 0.604 & 0.726   & 0.692    \\
\hline
\end{tabular}
\end{table}

To evaluate if the quantization of the models does not negatively impact the quality of the ranking we also evaluate the NDCG@10 scores of the quantized models on both BrightFit datasets by re-ranking the top 50 candidates with the default descriptions. The results for this are in Table~\ref{tab:res-exp5-quant-scores}. The dynamic quantization method achieves the best scores of the quantization methods, and both SmoothQuant and dynamic quantization lead to results that are not significantly lower than the FP32 model. The static quantization method causes a notable and significant decrease in performance. %It also did not achieve the best throughput in the first part of this experiment giving us little reason to use this quantization method going forward. 
%Overall we think that applying dynamic quantization to our models when using them in a production environment is a good way to speed up inference. Dynamic quantization also shows little to no degradation in performance and in some cases even improved performance.

\subsection{Experiment 5: User evaluation}\label{sec:exp6-user-exp}
%In this experiment, we aim to answer our final research question \textbf{[RQ5]}: \textit{"How do improvements on benchmark datasets translate to the user's perception of the recommendations?"} 
As per the recommendation by Knijenburg and Willemsen~\cite{Knijnenburg2015} we perform both an A/B test and a questionnaire to effectively evaluate the user perception of our course ranking method.

\subsubsection{A/B test}\label{sec:exp6a-ab-test}
We implement an A/B test in the BrightFit production environment. We use this test to directly compare the in-production system (BM25-based) against our best ranking setup.

\paragraph{Ranking model selected for test}
%Based on insights from the previous experiments,
We select a version of our RankT5 model that balances speed and quality: %It is important to have a good balance between these two aspects to ensure a good user experience. 
%We use the 
course descriptions summarized by the LongT5 model; %for re-ranking, because it is much faster to compute summaries for all courses with LongT5 than with Vicuna. So while the Vicuna summaries showed better performance on the general skills dataset, it is more practical for setting up the experiment in a timely manner to use the LongT5 summaries. %In experiment 2 we have seen that including the skill descriptions in the re-ranking query has a negative effect on performance so we do not include those. 
%Since we found in Section~\ref{sec:expquant} that the 
dynamic quantization; %method is able to achieve a nice speed-up for the models while sometimes even improving re-ranking performance, we select this method to use in the A/B test. We use 
a maximum input length of 256; %, as this has shown good performance in combination with the LongT5 summaries. %There is no real benefit of increasing the maximum input length to 512 as the summaries have a maximum length of 256. 
%Lastly, we use 
and a re-ranking depth of 20. In Section~\ref{sec:reranking} we have seen that a re-ranking depth of 30 achieves a higher NDCG@10 score. However, by reducing the number of processed documents from 30 to 20 we get close to a 150\% speed-up. We are still happy with the quality of the recommendations at a depth of 20, which is why we decide to use a depth of 20 instead of 30.

%When performing an A/B test there are two main ways of setting up how we randomly serve recommendations from system A or system B. These two randomization methods are between-subject randomization and within-subject randomization~\cite{Knijnenburg2015}. In a between-subject recommendation approach, a single user will only be exposed to one of the two systems. This type of setup is a more realistic scenario since in a real system, the user will also only get recommendations from a single system. When using within-subject randomization, the recommendations will be served randomly by either system A or B each time a request for a recommendation is made. The advantage of this method is that the differences between participants (Eg. one user might be really interested in learning courses while others may not be as interested) have a smaller influence on the end result. For our A/B test,

\paragraph{Experiment design} We use a between-subject randomization method as suggested by~\citet{Knijnenburg2015}. %as it represents a more realistic scenario. %Additionally, it is easier to set up a between-subject randomization for the BrightFit application with regard to tracking user interactions. 
%To decide whether we serve a user the course recommendations from either system A or B we use the user id which is a simple number. 
If a user's id is even-numbered, we use the in-production (BM25-based) ranker to select the presented recommendations and if the user's id is an odd number we use our RankT5-based system described above. We deploy the A/B test in the United States and Australia. The United States is the largest market for BrightFit and Australia uses the exact same skills taxonomy and course catalogue as the United States version, allowing us to easily deploy it in Australia. % without any additional modifications.

To track user behaviour during the experiment we use Google Analytics.\footnote{\url{https://marketingplatform.google.com/about/analytics/}} % a tool to track clicks on web pages. 
We track the following three events related to the course recommendations in BrightFit:
\begin{enumerate}
    \item \textit{Open skill card}: %The first interaction we monitor is a user opening a skill card. 
    A skill card is a small component presented to the user on the self-assessment report page (see Section~\ref{sec:queries}), showing the recommended skills. The user can click a skill card to get more information on a skill. Opening a skill card will also show them the list of courses that are recommended for that skill. This is the starting point for the recommender.
    \item \textit{Open course card}: Once a user has opened a skill card they will be presented with some information about the skill and a list of courses that are recommended to learn that skill. Each course is shown as a small card with some basic information like the title, the level, the rating and a thumbnail image. When a user is interested in a course they can open this course card to get more information about the course. For this event, we track if a user clicks on the course card to learn more about a course.
    \item \textit{Go to course}: %The final event we track is if the user proceeds to the actual course from the course provider. 
    Once a user has opened a course card and they are presented with the full description of the course they will have the option to go to the actual course on the course provider's website, which is what we track with this event.
\end{enumerate}

We ran the experiment for 66 days in the United States and for 46 days in Australia.

% \begin{figure}[t]
%     \centering
%     \makebox[\textwidth][c]{\includegraphics[width=1.2\textwidth]{funnel-combined-user-unique.pdf}}%
%     \caption{Funnel diagram for the number of unique users that trigger events recorded during the BrightFit A/B test combined for the USA and Australia. The first value in each box indicates the total number of recorded events, the percentage indicates how many events of that type were recorded compared to the previous stage in the funnel.}
%     \label{fig:res-exp6-funnel-unique-user}
% \end{figure}

%In the first funnel diagram, we can see that we have a total of 361 times that a skill card has been opened. In both funnel diagrams, we see that the distribution between system A and system B is almost identical which means that we chose a good method of splitting the users between system A and system B.

\begin{figure*}[h]
    \centering
    \includegraphics[width=0.7\textwidth,trim={0cm 2cm 0 0},clip]{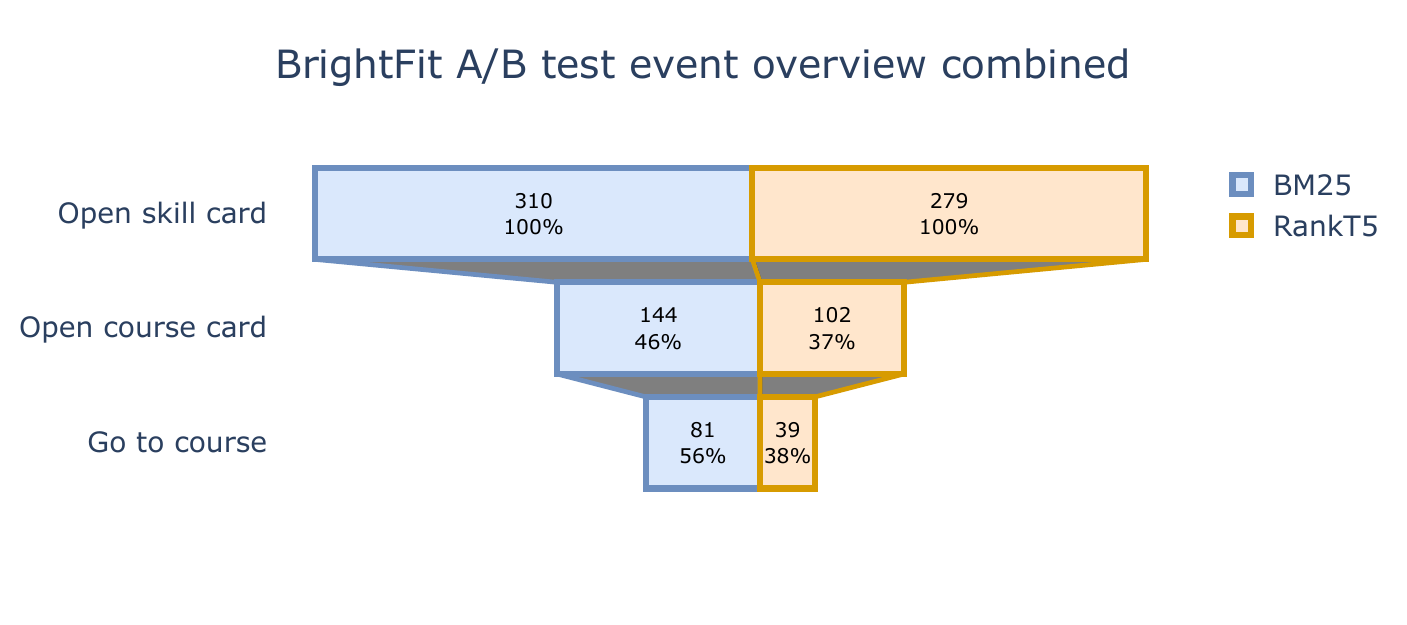}%
    \caption{Funnel diagram for the events recorded during the BrightFit A/B test combined for the USA and Australia. The first value in each box indicates the total number of recorded events, the percentage indicates how many events of that type were recorded compared to the previous stage in the funnel.}
    \label{fig:res-exp6-funnel}
\end{figure*}

\paragraph{Results} We show the combined results for both countries in Figure~\ref{fig:res-exp6-funnel}. %Additionally, we show the number of unique users that have triggered the events in Figure~\ref{fig:res-exp6-funnel}.
Overall we find that the BM25-based recommender achieves higher click-through rates between the different parts of the BrightFit application. In Figure~\ref{fig:res-exp6-funnel} we can see that, after a user opened a skill card and was shown course recommendations by the BM25-based recommender, they open a course card in 50\% of the cases, compared to only 29\% for the RankT5-based recommendations. This contrasts our offline evaluation results in Table~\ref{tab:res-exp-2}. We discuss this contrast in Section~\ref{sec:discussion}. % which is substantially lower. %From Figure~\ref{fig:res-exp6-funnel-unique-user} we can see that over 35\% (25/70) of the users that open at least one skill card will also end up clicking through to a full course from the external provider. This is a lot higher compared to the almost 24\% (17/71) of users in served recommendations by the RankT5 system.

\subsubsection{User questionnaire}\label{sec:exp6b-survey}
Again per the recommendation by~\citet{Knijnenburg2015} we use an additional user questionnaire to get more insight into user preferences. %what users actually look at and prefer when getting recommendations. 

\begin{figure*}[t]
    \centering
    \includegraphics[width=\textwidth]{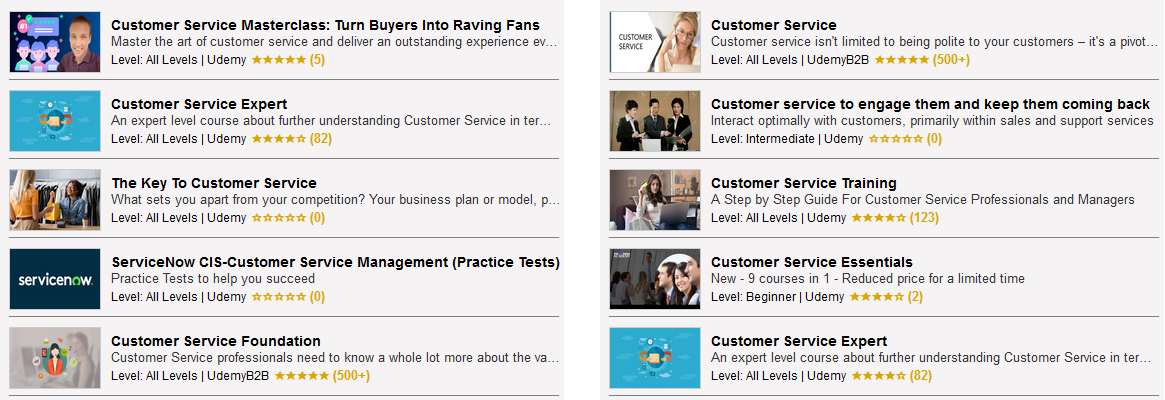}%
    \caption{An example of the two lists a user is shown for the query \textit{Customer Service for Front Desk Employee} in the questionnaire.}
    \label{fig:survey-example}
\end{figure*}

\paragraph{Study design} We present users with the 10 queries from the BrightFit general skills dataset with their top recommended courses. We have to be careful about how we present the recommendations as this can have a significant influence on the user's choices~\cite{Knijnenburg2015}. We also do not want to overwhelm the users with choices to prevent so-called ``choice overload"~\cite{Bollen2010choice-overload}. After careful consideration and discussing options with test users, we opted for 5 courses per list. We show the two lists below each other for mobile-friendliness and we %The layout of the recommendations also plays an important role. When the recommendations are shown in a vertical list, users pay a lot more attention to the first handful of items in the list~\cite{Bollen2010choice-overload}. % We also have to decide if we show the two generated recommendation lists side-by-side or below each other. Chen et al.~\cite{chen2011recommender-interface} discuss that when recommendations are divided over two separate pages, the users rarely pay attention to the recommendations on the second page. To eliminate this effect as much as possible and 
%To reduce user bias, 
randomize which system's recommendations are first~\cite{chen2011recommender-interface}. %We also have to be wary of this when we decide how to lay out the questions in the questionnaire. 

To make the questionnaire as representative as possible of the way recommendations are served to the user in BrightFit, we show them only information that is also available in BrightFit: %. When a user gets shown the list of course recommendations, the main things they get to see are 
the course title, course thumbnail, course rating, course provider, and the course level. An example of two lists we generate for a query in the questionnaire is shown in Figure~\ref{fig:survey-example}. Note that in this figure, the lists are laid out side by side while the lists in the survey are presented below each other.

For each of the 10 questions, we ask the user to indicate which list of recommendations they prefer or if they have no preference for one list over the other. We then ask them to elaborate on their reasoning for the choice they made to get an understanding of what users find important when getting recommendations. This has the additional benefit that it improves the accuracy of the responses given, as the users are forced to think more about their answer~\cite{Knijnenburg2015}. %All 10 queries that we generate recommendations for in the questionnaire are listed in our github repository.

\begin{figure}[t]
    \centering
    \includegraphics[width=0.9\columnwidth,trim={1.6cm 0 2.5cm 0},clip]{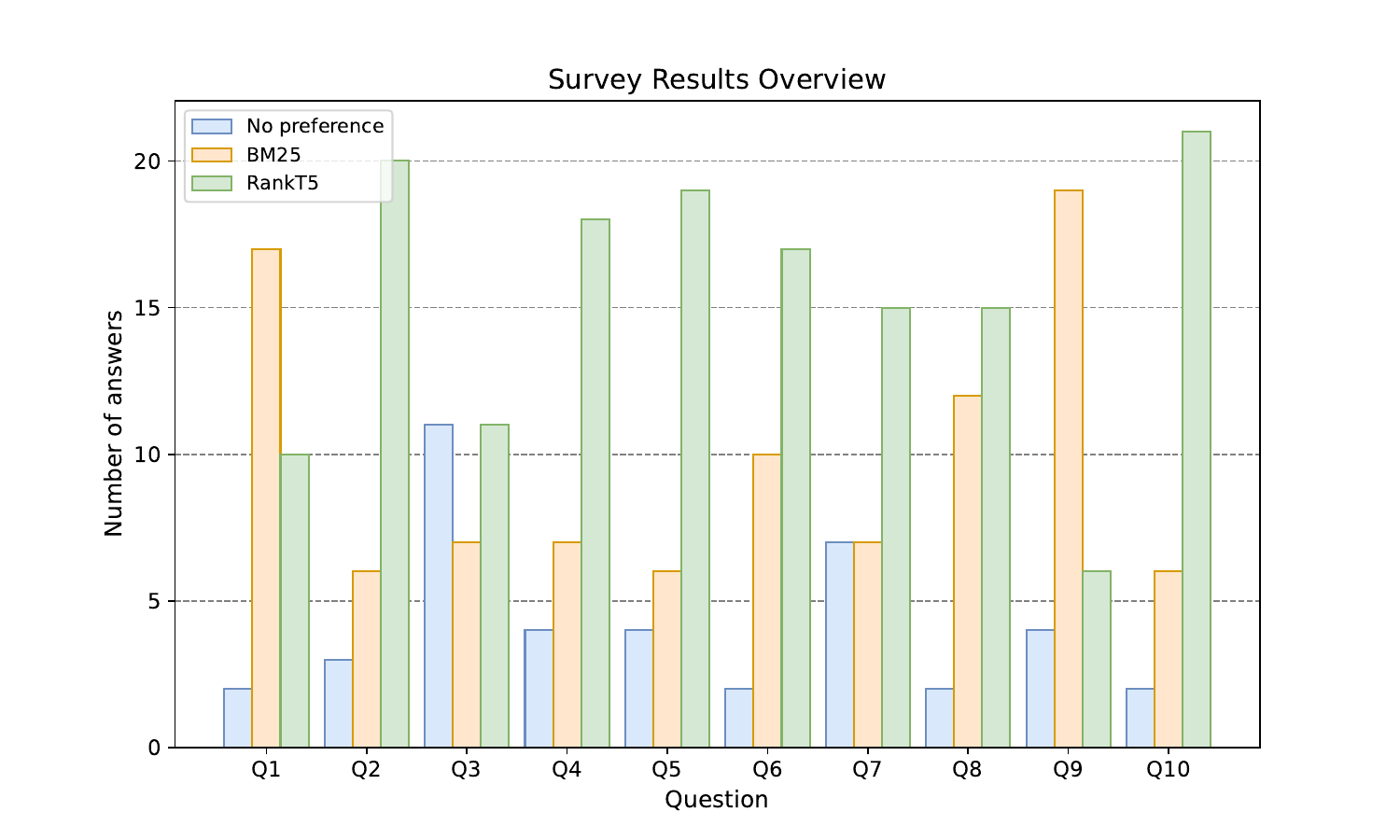}
    \caption{Results of the conducted user questionnaire showing user preference for each of the 10 questions in the questionnaire. In total responses from 29 users were collected.}
    \label{fig:res-exp7-preference}
\end{figure}

\paragraph{Results} In total, we have collected responses from 29 different participants (M:19, F:9, X:1; mean age:28). Figure~\ref{fig:res-exp7-preference} shows the user preferences for each question in the questionnaire. We can see that in 8 out of 10 cases, the results from the two-stage retrieval approach using GTR retrieval and RankT5 re-ranking are preferred over the BM25-based in-production ranker. This result is in contrast with the results we observed in the A/B test, but aligns with our offline evaluation results. We will discuss this in Section~\ref{sec:discussion}.

Even though the new recommendation system gets preferred over the old recommendation system in the questionnaire, we still have the case that for each query there is at least a handful of users that have no preference or prefer the other system. This shows that there is a difference between users' perceptions of what makes a good recommendation. We looked at the motivations written by the users for the choices they made. As reasons were mentioned: higher course ratings, courses for different levels, diversity in course content, false positives, and courses taught within the job context.

\section{Discussion}
\label{sec:discussion}
In this section, we further discuss the results of the experiments, focussing on the unexpected findings. %We perform additional analysis and look at deeper interpretations of the results. We also discuss some limitations of our work.

\paragraph{Comparing datasets} %If we take a general look at the results of the experiments where we evaluate methods on both the BrightFit datasets, 
We found that the scores on the general skills dataset are generally higher than the scores on the BrightFit IT dataset. This shows us that the BrightFit IT dataset is a harder problem compared to the general skills many more courses for IT-related skills available, which can make it harder to get a good ranking of these courses. This is also supported by the fact that the BrightFit IT skills dataset has more relevant documents per query compared to the BrightFit general skills dataset. % as shown in Tables~\ref{tab:bf-it-overview} and \ref{tab:bf-general-overview}. 

%\paragraph{Re-ranking depth} We found that the best re-ranking depth depends on the dataset. %Every time we increase the re-ranking depth we see that the recall increases, meaning that there are more relevant documents in the candidate set. However, we also add irrelevant documents to the candidate list when we increase the re-ranking depth. 
%For some datasets, we see an increase in NDCG@10 scores when increasing the depth. This is likely the case when the documents in the datasets are not that similar, which makes it easier for the re-ranker to create a good ordering. In cases where the documents in a dataset are very similar to each other, it can be harder for the re-ranker to create a good ordering when there are more documents in the candidate set. As we have seen it is very dataset-dependent on what depth works best, showing that it is important to tune the re-ranking depth. 

\paragraph{Quantization} %We found that the SmoothQuant quantized model caused a significant performance drop in terms of NDCG@10 scores on both the BrightFit datasets, even when compared to the dynamic quantization method. We believe that this is likely caused by the value of the hyperparameter $\alpha$ that we set for the SmoothQuant algorithm. A higher value of $\alpha$ is more suitable for models that have activations with a large number of outliers, while a lower $\alpha$ is more suitable for models with fewer outliers in the activations. Since our model is relatively small and thus has fewer outliers we should be able to get better performance with a lower value of $\alpha$. 
We found the dynamically quantized model performs very similarly to the FP32 movel, and in some settings even a bit better. This latter finding is surprising since quantization makes the model weights less precise, causing the model to have less information. We speculate that the model might be able to generalize better with the less specific weights after quantization, resulting in better scores.

\paragraph{User studies} In the %final experiments, we evaluated our methods using real users by performing both an A/B test as well as creating a questionnaire to survey users.
user evaluation, the results from the A/B test and the questionnaire contradict each other. The result of the A/B test shows better user interaction with the BM25-based recommender, while the users who have taken the questionnaire have clearly shown a preference for the RankT5-based recommendations, the latter confirming our offline ranking experiments. %While the A/B test is more representative of the BrightFit application compared to the questionnaire, there are some factors in the A/B test that we do not measure or look at, which can be the reason for the unexpected results. 
One of the factors in the A/B test might be the interpretability of the results. If we have the query \textit{``Information Security for Security Specialist''} the BM25-based re-ranker will retrieve course titles that contain the words ``Information Security". On the other hand, we observe that in the RankT5-based recommendations there are many courses with the term \textit{CISSP} which stands for ``Certified Information Systems Security Professional". For someone who does not know \textit{CISSP} stands for, % which can often be the case as people in BrightFit are looking to learn new skills, 
the recognizable course descriptions retrieved by BM25 might be more appealing to click on. 

Another factor that can be an issue is the time it takes to generate recommendations, which is a relevant aspect given that efficiency was one of the aims of our approach. The BM25-based recommendations are much faster to generate than the recommendations by RankT5. During our testing of the A/B test setup, we found that in almost all cases, the recommendations for the courses by RankT5 would be done by the time a normal user would get to the part of the page where the skills are shown, which is near the bottom. However, despite our efforts to improve the inference time using quantization and summarization, we cannot be sure that this longer time to generate the recommendations is impacting how the users interact with the system. This difference in time is not a factor in the questionnaire, which can explain the difference in results between the A/B test and the questionnaire results.

\section{Conclusion}\label{sec:conclusion}
In this paper, we addressed the problem of making good and efficient course recommendations for users who are looking to learn a certain skill. We evaluated a two-stage retrieval approach using GTR and RankT5 on two newly labelled datasets based on the commercial BrightFit course recommender. We experimented with improving the efficiency of the recommender by applying summarization and quantization techniques. %Additionally, we looked at re-ranking depth and quantization methods to speed up our model and make it more suitable for use in the BrightFit application. Lastly, we performed multiple experiments to evaluate our methods and answer the following research questions:

%\paragraph*{[RQ1]} \textbf{How suitable are transformer-based first-stage retrievers for our use case?} 
We found that the zero-shot GTR retriever outperforms the BM25-based retriever on both BrightFit datasets. %On the IT dataset, it was able to make a substantial improvement on the recall@20 score from $0.380$ to $0.512$. On the BrightFit general skills dataset, it was able to improve the recall@20 score from $0.400$ to $0.545$ compared to the BM25-based method. %This has shown us that the GTR retriever is a good choice for our use case of recommending courses. 
%\paragraph*{[RQ2]} \textbf{How effective are zero-shot re-ranking methods for ranking course recommendations?} In this research, we have looked at applying the RankT5 model as a zero-shot re-ranker for recommending BrightFit courses. 
We finetuned the RankT5 model on the MSMARCO dataset and %then looked at different variations of this trained model for applying it to the BrightFit data. We have seen that we are able to improve the quality recommendations by re-ranking the courses that are retrieved by the GTR model. This 
showed that it can effectively perform course re-ranking despite not being trained on the course domain, and leads to a substantially better ranking for two labelled datasets than the in-production BM25 ranker.
%\paragraph*{[RQ3]} \textbf{Can we improve ranking performance by applying summarization techniques to the course descriptions?} 
We compared two summarization models for improving ranking of long course descriptions: a LongT5 model fine-tuned on edX data for summarizing courses and a Vicuna model that we prompted to generate summaries. %While the LongT5 summarizer performed better on the BrightFit IT dataset and the Vicuna summarizer performed better on the BrightFit general skills dataset, 
We found that both models were able to get better performance than was achieved with the original course descriptions. %This shows us that summarization techniques are an effective method of improving ranking performance for the BrightFit course descriptions.
% \paragraph*{[RQ4]} \textbf{What is the impact of changing the re-ranking depth on the ranking performance?}\\
% For both the BrightFit datasets as well as the BEIR benchmark we looked at different re-ranking depths and how they affect the NDCG@10 scores. We noticed a lot of variation in the results for the BEIR datasets where some datasets would benefit from a lower re-ranking depth, while others performed better with a higher re-ranking depth. This shows us that the optimal re-ranking depth is dataset dependent and we cannot draw one general conclusion regarding the impact of the re-ranking depth. On average, the difference between the best and worst re-ranking depth for the datasets in the BEIR benchmark was around $5.5\%$ with the re-ranking depth of 20 performing the best on average. The largest difference we observed between the best and worst-performing re-ranking depths was on the ArguAna dataset with a difference of over $48\%$. For some use cases, as we have with BrightFit, we also have to take into account the additional time it takes to compute the recommendations when using a higher re-ranking depth. This makes it even more difficult to draw a strong conclusion about what an optimal re-ranking depth is.
%\paragraph*{[RQ4]} \textbf{What effect do different quantization methods have on the inference speed and prediction quality?} As part of our research, 
We investigated the use of quantization to speed up model inference % We have looked at 3 quantization methods: static quantization, dynamic quantization and SmoothQuant. In the experiments, we have seen
%and found that dynamic quantization offers the best performance and inference speed out of these three quantization methods. %The static and SmoothQuant quantization methods were both slower and had a significant decrease in prediction quality compared to the dynamic quantization. 
%For the dynamic quantization, we observed 
and we achieved a speed-up of up to 47\% %for the 256 input length 
compared to inference with the original (FP32) model, while the performance was either very close or sometimes even better than the FP32 model. %We think that applying dynamic quantization to the model is a good way of obtaining a nice speed-up without impacting the quality of the predictions negatively.

%\paragraph*{[RQ5]} \textbf{How do improvements on benchmark datasets translate to the user's perception of the recommendations?} To answer this research question 
Finally, we performed 2 user studies: an A/B test and a questionnaire. From the results of the A/B test, we saw that the BM25-based re-ranker had substantially higher click-through rates than our proposed two-step approach, %with 50\% of the users clicking through from a skill card to a course card compared to only 29\% for the RankT5-based recommendations. This finding 
which does not align with our findings that RankT5 was significantly better on both BrightFit benchmark datasets. Time to generate recommendations and transparency of relevance might be factors that play a role here. %We discussed some reasons for the unexpected observations but more research is needed to truly find out what the cause of the large difference is.
%In addition to the A/B test, we also surveyed users to get more explicit feedback on the recommendations by the BM25- and RankT5-based recommenders. In this 
With the questionnaire, on the other hand, we found that in 8 out of 10 cases, the RankT5-based recommender was either equal to or preferred over the BM25-based recommender. %This is in contrast to the findings of the A/B test and this result aligns more with our findings on the benchmark dataset. 
Through the questionnaire, we gained insights into why users preferred one set of recommendations over another which is useful information when looking to extend the recommender in the future.

%Overall we think that more investigation is required into the results of the A/B test before we can fully draw a conclusion for this research question, but based on the explicit feedback of the questionnaire we can say that the changes we made to the recommendations have a positive effect on the users' perception of the recommendations. 

%Overall the best NDCG@10 score we obtained on the BrightFit IT dataset is a score of $0.684$ with the RankT5 model using the LongT5 generated course descriptions and a maximum input length of 256 without including the skill descriptions in the query. For the BrightFit general skills dataset, the best nDCG score we obtain is $0.844$ with the Vicuna generated summaries, a 256 maximum input length and not also not using the skill descriptions in the query. These scores show a good improvement over the $0.482$ and $0.447$ NDCG@10 scores the BM25-based recommender obtains on the BrightFit IT and general skills datasets respectively.

%Future work to improve and extend the recommender system would be to look at the user feedback we have gathered through the questionnaire. We have identified four main aspects that users value a lot, but we have not actively modelled any of these in our methods. By looking at incorporating one or multiple of these four key aspects we believe that the recommendations can be improved even further.
Future work could focus on differences between the preferences of individual users. An extension incorporating user preferences might provide a better user experience by being able to serve personalized recommendations.
Another way to improve recommendations would be to construct a proper training dataset for BrightFit with enough ratings so that we can directly train on the task instead of using zero-shot methods.
%% Define the bibliography file to be used
\bibliography{thijmen,added}


\end{document}